\documentclass[conference]{IEEEtran}

\usepackage{amssymb}
\usepackage[cmex10]{amsmath} 
\usepackage{amsthm}  
\usepackage{amsfonts}
\usepackage{pgf}
\usepackage{pgffor}
\usepackage{tikz}
\usepackage{mathtools} 
\usepackage{pst-sigsys}
\usepackage{stfloats}
\usepackage{graphicx}
\usepackage{cite}
\usetikzlibrary{patterns}
\usepackage{subfigure}

\newcommand{\comment}[1]{}
\newcommand{\varx}{\sigma^2_X} 
\newcommand{\varE}{\sigma^2_E}

\newcommand{\dL}{d_L}
\newcommand{\dK}{d_K}

\newcommand{\eps}{\epsilon}

\newcommand{\Ostergaard}{\text{\O stergaard }}

\newcommand{\RKL}{R_{K,L}}

\newcommand{\logtwo}[1]{\log_2{\left({#1}\right)}}

\newcommand{\EntRate}[1]{\bar{H}\left(#1\right)}
\newcommand{\DifEnt}[1]{h\left(#1\right)}
\newcommand{\DifEntRate}[1]{\bar{h}\left(#1\right)}

\newcommand{\sinc}[1]{\ensuremath{\text{sinc}}\left({#1}\right)}
\newcommand{\refequ}[1]{(\ref{#1})}
\newcommand{\reffig}[1]{Fig.~\ref{#1}}
\newcommand{\reftheo}[1]{Theorem~\ref{#1}}
\newcommand{\reflem}[1]{Lemma~\ref{#1}}

\newcommand{\str}[2]{{#1}^{(#2)}}

\newcommand{\lat}{\Lambda}
\newcommand{\var}[1]{\sigma_{#1}^2}

\newcommand{\size}[1]{|#1|}

\newcommand{\ith}{$i^{\text{th}}$ }
\newcommand{\osr}{\gamma}

\newcommand{\vecstyle}[1]{\boldsymbol{#1}} 
\newcommand{\seqstyle}[1]{\vecstyle{#1}} 
\newcommand{\vect}[1]{\vecstyle{#1}}  
\newcommand{\rvec}[1]{\vect{\rvar{#1}}}  
\newcommand{\rseq}[1]{\rvec{#1}}  
\newcommand{\seq}[1]{\vect{#1}}
\newcommand{\rvar}[1]{\uppercase{#1}} 

\newcommand{\set}[1]{{#1}}
\newtheorem{theorem}{Theorem}
\newtheorem{lemma}{Lemma}
\newtheorem{remark}{Remark}

\begin{document}
\sloppy

\title{Sampling versus Random Binning for Multiple Descriptions of a Bandlimited Source}

\author{
  \IEEEauthorblockN{Adam Mashiach}
  \IEEEauthorblockA{Dept. Electrical Engineering-Systems\\
    Tel-Aviv University\\
    Tel-Aviv, Israel\\
    Email: adam.mashiach@gmail.com} 
  \and
  \IEEEauthorblockN{Jan  \Ostergaard}
  \IEEEauthorblockA{Dept. Electronic Systems \\
    Aalborg University\\
	 Aalborg, Denmark\\
    Email: jo@es.aau.dk}
  \and
  \IEEEauthorblockN{Ram Zamir}
  \IEEEauthorblockA{Dept. Electrical Engineering-Systems\\
    Tel-Aviv University\\
    Tel-Aviv, Israel\\
    Email: zamir@eng.tau.ac.il}
}
\maketitle

\begin {abstract}
Random binning is an efficient, yet complex, coding technique for the symmetric $L$-description source coding problem. We propose an alternative approach, that uses the quantized samples of a bandlimited source as ``descriptions''. By the Nyquist condition, the source can be reconstructed if enough samples are received. We examine a coding scheme that combines sampling and noise-shaped quantization for a scenario in which only $K<L$ descriptions or all $L$ descriptions are received. Some of the received $K$-sets of descriptions correspond to uniform sampling while others to non-uniform sampling. This scheme achieves the optimum rate-distortion performance for uniform-sampling $K$-sets, but suffers noise amplification for nonuniform-sampling $K$-sets. We then show that by increasing the sampling rate and adding a random-binning stage, the optimal operation point is achieved for any $K$-set.
\end {abstract}


\section{Introduction}
\label{sec:Into}
Sampling can be viewed as a signal-processing analogue of random binning. When using random binning, lossless reconstruction is possible as long as the binning rate is higher than the source entropy, regardless of the source specific distribution. Similarly, perfect reconstruction of a bandlimited signal is possible when the average sampling rate is at least twice the signal bandwidth (Nyquist rate), regardless of its specific spectrum. Furthermore, just as samples can be accumulated at arbitrary time instances, partial binning information can be combined, until the condition for perfect (lossless) reconstruction is met. These universality properties extend to a vector of correlated sources by the Slepian-Wolf-Cover theorem for random binning, and by the vector sampling expansions theorems for sampling.

We study the potential and limitations of this analogy in the multiple description (MD) problem. Multiple descriptions is a joint source-channel coding problem, in which several ($L$) coded representations (descriptions) of the source are created. The source can be reconstructed from any subset of received descriptions, with resulting distortion that decreases with the number of received descriptions. While the focus in the past was mainly on the two-description case (e.g., \cite{Cover&ElGamal},\cite{Ozarow}), the many-descriptions ($L>2$ descriptions) case is recently getting more attention, as a good framework for robust multimedia transmission over packet-switching networks in the presence of packet loss. 

The Gaussian MD rate-distortion region is not known for $L>2$ descriptions, and most research focus on certain special cases. An interesting special case is the symmetric MD, in which all the descriptions have the same rate, and the distortion depends only on the number of received descriptions. The best known achievable schemes for the symmetric MD problem (Gaussian source and MSE) are based on a coding scheme that was proposed by Puri, Pradhan and Ramchandran (PPR) in \cite{Pradhan&Puri&Ramchandran1}. The key concept of the PPR scheme is "randomly binned codebooks", which is inspired by source coding with side information. It enables the encoder to encode each description while treating the other $L-1$ as potential side information, which may be available at the decoder, and thereby reduces the coding rate. While it is unknown whether this scheme is optimal for the general (Gaussian) symmetric case, it is optimal for a special case, in which one is interested only in receiving some $K < L$ descriptions or all $L$ descriptions \cite{Wang&Viswanath:2009}. We refer to this special case as the ``$K$-or-$L$`` problem. 

From a practical point of view, however, there is a need for a coding scheme that can easily generate a large number of descriptions, while not sacrificing too much in performance. One such a coding scheme was presented by the authors in \cite{Mashiach&Ostergaard&Zamir}, where the two-description solution of \cite{Ostergaard&Zamir} is extended to $L$ descriptions, and is proved to be optimal for the $1$-or-$L$ MD problem ($K=1$). This scheme is based on oversampling and dithered lattice quantization with noise shaping, and is referred to as the DSQ scheme, since it was inspired by delta-sigma quantization. In the DSQ scheme each description consists of quantized source samples taken at different time instances. In \cite{Mashiach&Ostergaard&Zamir} each description was sampled exactly at the source's Nyquist rate, thus, the source could be reconstructed from any single description. As more descriptions are received, the decoder can use the fact that the source is bandlimited, to filter some of the quantization noise, and thus reduce the distortion. The noise shaping operation enables to trade-off the side distortion (single description) for the amount of improvement in distortion with any additional received description.

In this paper, motivated by the analogy between random binning and sampling, we study two coding schemes based on the DSQ scheme for the $K$-or-$L$ problem. In the first scheme, each description is sampled at $1/K$ of the source's Nyquist rate, and is referred to as DSQ with sub-Nyquist sampling scheme. In this case, reconstruction without aliasing is possible when receiving $K$ descriptions or more. We show that while this solution achieves the optimal performance for received $K$-description sets that correspond to a uniform sampling pattern, the other sets suffer from higher distortion due to noise amplification in nonuniform sampling \cite{Mashiach&Zamir}. In the second scheme, to avoid noise amplification, we sample each description at the source's Nyquist rate as in \cite{Mashiach&Ostergaard&Zamir} and use random binning coding to compensate for the redundancy due to the oversampling (each $K$ descriptions are sampled $K$ times faster than the Nyquist rate)\comment{oversampled $K$ descriptions}. We prove that this scheme achieves the same performance as the PPR scheme; thus, it is optimal for the $K$-or-$L$ problem.

The paper is organized as follows. Section~\ref{sec:background} formulates the $K$-or-$L$ problem and presents the PPR and DSQ schemes. In Section \ref{sec:NoRB} we study the DSQ scheme with sub-Nyquist sampling, while in Section~\ref{sec:DSQ-RB} we combine the Nyquist DSQ scheme with random binning. Section~\ref{sec:conclusion} concludes the paper.

\section{The PPR and DSQ Coding Schemes}
\label{sec:background}
We begin this section with some notation\comment{ that will be used throughout the paper}. We use upper case letters ($\rvar{x}$) for stochastic variables and lower case letters ($x$) for their realization. Vectors or infinite sequences will be indicated by bold face ($\rvec{x}$). For any sequence $\seq{a}$ we define its \ith \emph{stream} by the subsequence $a^{(i)}_n=a_{i+nL}$, where $L$ is the number of descriptions. For any set of indices $\set{J} \subseteq \left\{0,1,...,L-1\right\}$ we define $\seq{a}^{(J)}$ as the vector process $\seq{a}^{(J)}_n=(a_n^{(j_1)},a_n^{(j_2)},...,a_n^{(j_{\size{\set{J}}})})$.

Now we will briefly present the $K$-or-$L$ MD problem (for a more complete formulation see \cite{Wang&Viswanath:2009}). Let $\seqstyle{X}$ be a stationary and memoryless Gaussian process with zero mean and variance $\varx$. $\seqstyle{X}$ is encoded by $L$ encoding functions to produce $L$ descriptions at equal rate $R$. Denote the distortion (MSE) achieved at the decoder when receiving a set $\set{J}$ of descriptions by $d_{\set{J}}$. In the symmetric MD problem the distortion $d_{\set{J}}$ depends on $\set{J}$ only through $\size{\set{J}}$, thus we can replace $d_{\set{J}}$ by $d_{\size{\set{J}}}$. In the $K$-or-$L$ problem only distortion constraints $\dK$ and $\dL$ are considered. The minimum achievable rate for given distortion constraints $(\dK,\dL)$ is the rate-distortion function (RDF) and is denoted by $\RKL(\dK,\dL)$. A common representation of the MD problem is of many receivers, where each receives a different set of descriptions. In the $K$-or-$L$ problem there are $\binom{L}{K}$ "first layer" receivers, where each receives a different set of $K$ descriptions, and a "central receiver" which receives all $L$ descriptions. 

In \cite{Wang&Viswanath:2009} Wang and Viswanath gave $\RKL(\dK,\dL)$ implicitly as an optimization problem for a vector Gaussian source with MSE fidelity criterion. For a scalar Gaussian source, the explicit solution to this optimization problem is (see \cite{Mashiach&Ostergaard&Zamir})
\begin{align}
\RKL(\dK,\dL) &= \frac{1}{2K}\logtwo{ \frac{(L-K)(\varx-\dL)}{L(\dK-\dL)}} 
\nonumber \\
&\quad+\frac{1}{2L}\logtwo{ \frac{K\varx(\dK-\dL)}{(L-K)\dL(\varx-\dK)}}
.
\label{equ:RKL}
\end{align}
We observe that the total rate $L\RKL(\dK,\dL)$ depends on $(K,L)$ only through the ratio $L/K$.\comment{ Thus, if $K$ divides $L$, we have $L\RKL(\dK,\dL)=L/K\cdot R_{1,L/K}(d_1,d_{L/K})$ when substituting $d_1=\dK$ and $d_{L/K}=\dL$. }
In other words, the RDF for the $K$-or-$L$ problem equals the RDF for the $1$-or-$L/K$ problem. We refer to this as the ``scaling property'' of the RDF and exploit it in Section~\ref{sec:NoRB}. 

\textbf{PPR scheme:}\comment{Since we use the PPR coding scheme as a baseline for comparison, we will briefly introduce it for a Gaussian source.} In \cite{Pradhan&Puri&Ramchandran1}, Pradhan et al gave a new coding scheme for the general symmetric MD problem, to which they referred as $(L,K)$ source-channel erasure codes, and we refer to it simply as the PPR scheme. Although in \cite{Pradhan&Puri&Ramchandran2} they extend this scheme by layering several such codes with different $K$ values, in this paper we consider only one layer (as in \cite{Pradhan&Puri&Ramchandran1}). 

The PPR coding scheme consists of two steps as follows. In the first step the source is encoded using $L$ independent Gaussian codebooks with rate $R'$. The \ith codebook is constructed using the marginal distribution of the random variables $\left\{\rvar{y}_i\right\}_1^L$ given by $\rvar{y}_i=\rvar{x}+\rvar{v}_i$, where $\left\{\rvar{v}_i\right\}_1^L$ (denoted by $\rvar{Q}_i$ in \cite{Pradhan&Puri&Ramchandran1}) are identically distributed jointly Gaussian random variables (independent of $\rvar{x}$) with variance $\var{V}$ and pairwise correlation coefficient $\rho$. The codewords in each codebook are randomly assigned to $2^{nR}$ bins, and in the second coding step each\comment{ chosen codeword from the first step} codeword is encoded using its bin index. The PPR binning rate $R$ should be high enough so that when receiving some $\size{\set{J}} \geq K$ descriptions the decoder can find only one $\size{\set{J}}$-tuple of codewords, one from each relevant codebook, that are jointly typical. The random binning coding is the key component of the PPR scheme, and it enables the encoder to use the fact that at least $K$ descriptions are available at the decoder in order to reduce the coding rate. We notice that while the first coding step is lossy, the second one is lossless (when receiving at least $K$ descriptions).

The distortions and rate of the PPR scheme (for $\varx=1$) are given in \cite{Pradhan&Puri&Ramchandran1} by
\begin{equation}
d_{\set{J}}=\frac{\var{V}\left[ 1+(\size{\set{J}}-1)\rho \right]}{\size{\set{J}}+\var{V}\left[ 1+(\size{\set{J}}-1)\rho \right]}, \quad \forall \set{J}: \size{\set{J}} \geq K
\label{equ:PPR_Dj}
\end{equation}
\begin{equation}
R=\frac{1}{2}\log_2{\!\!\left[\left[\frac{K+\var{V}\left(1+(K-1)\rho\right)}{\var{V}(1-\rho)}\right]^{\frac{1}{K}}
\!\left[\frac{1-\rho}{1+(L-1)\rho}\right]^{\frac{1}{L}}\right]}
\label{equ:R_PPR}
\end{equation}
The PPR scheme was proved to be optimal for the $K$-or-$L$ problem for a memoryless Gaussian source by Wang and Viswanath in~\cite{Wang&Viswanath:2009}. Thus, if we express $\rho$ and $\var{V}$ as a function of $d_L$ and $d_K$ by using \refequ{equ:PPR_Dj}, then \refequ{equ:R_PPR} coincides with \refequ{equ:RKL}.

\begin{figure*}
\centering
\begin{pspicture}[showgrid=false](0,-1.5)(10.25,2.5)

\psset{linewidth=0.5pt}
\dotnode[dotstyle=square*,dotscale=0.001](0,0){intersection0}
\psblock(1.5,0){upSampling}{ ${\uparrow}\osr$ }
\pscircleop(3,0){oplus1}  
\dotnode[dotstyle=square*,dotscale=0.001](5.3,0){intersection1}
\pscircleop(5.8,0){oplus3} 
\psblock(7,0){quant}{ $Q_M$ }
\dotnode[dotstyle=square*,dotscale=0.001](7.8,0){intersection2}
\pscircleop(5.3,-0.8){oplus2} 
\psblock(3.8,-0.8){filt}{$c'(z)$}
\dotnode[dotscale=0.7](8,0){dotCenter}

\newcount\cnt
\cnt=0
\psforeach{\ry}{1,0.2}{\advance\cnt by 1\relax
\dotnode[dotscale=0.7](8.5,\ry){dot\the\cnt}
\psblock(9.5,\ry){EC\the\cnt}{LC}
\dotnode[dotscale=0.001](10.25,\ry){des\the\cnt}
}

\dotnode[dotscale=0.7](8.25,-0.8){dotL}
\psblock(9.25,-0.8){ECL}{LC}
\dotnode[dotscale=0.001](10,-0.8){desL}
\psarc[style=Arrow]{<-}(7.6,0){0.5}{0}{60}

\pssignal[signalsep=0.1](9.25,2){ditherE}{Dither ($\seq{z}$)}
\pssignal[signalsep=0.05](5.05,-1.6){ditherDSQ2}{$z_k$}
\pssignal[signalsep=0.05](5.55,1){ditherDSQ1}{$z_k$}


\ldotsnode[angle=90](9.2,-0.3){dots}

\psset{style=Arrow}
\ncline{intersection0}{upSampling} \naput{$x_n$}
\ncline{upSampling}{oplus1} \naput{$a_k$}
\ncline{oplus1}{oplus3} \naput{$a'_k$}
\ncline{oplus3}{quant}
\ncangle[angleA=270,angleB=0]{intersection2}{oplus2}
\ncline{intersection1}{oplus2} \nbput[labelsep=0.1,npos=0.8]{--}
\ncline{oplus2}{filt} \naput{$e_k$}
\ncline{ditherDSQ2}{oplus2}\nbput[labelsep=-0.3,npos=0.6]{--}
\ncline{ditherDSQ1}{oplus3}
\ncangle[angleA=180,angleB=270]{filt}{oplus1} \naput{$\tilde{e}_k$}
\ncline{-}{quant}{dotCenter} \naput{$a_{q,k}$}
\ncline{dotCenter}{dot1}
\ncline{ditherE}{EC1}
\ncline{dot1}{EC1} \naput{$a^{(0)}_{q,n}$}
\ncline{dot2}{EC2} \naput{$a^{(1)}_{q,n}$}
\ncline{dotL}{ECL} \naput{$a^{(L-1)}_{q,n}$}
\ncline[doubleline=true]{EC1}{des1} \naput{$R$}
\ncline[doubleline=true]{EC2}{des2} \naput{$R$}
\ncline[doubleline=true]{ECL}{desL} \naput{$R$}
\end{pspicture}
\begin{pspicture}[showgrid=false](0.25,-1.5)(7.25,1.6)
\psset{linewidth=0.5pt}

\newcount\cnt
\cnt=-1
\psforeach{\ry}{1,0.2}{\advance\cnt by 1\relax
\dotnode[dotscale=0.001](0.25,\ry){in\the\cnt}
\dotnode[dotscale=0.001](0.5,\ry){EDin\the\cnt}
\dotnode[dotscale=0.001](1.25,\ry){EDout\the\cnt}
\dotnode[dotscale=0.001](2.5,\ry){stream\the\cnt}
}
\dotnode[dotscale=0.001](0,-0.8){inL}
\dotnode[dotscale=0.001](0.25,-0.8){EDinL}
\dotnode[dotscale=0.001](1,-0.8){EDoutL}
\dotnode[dotscale=0.001](2.25,-0.8){streamL}

\pssignal[signalsep=0.1](0.625,2){ditherD}{Dither ($\seq{z}$)}
\dotnode[dotscale=0.001](0.625,1.4){ditherCon}

\psfblock[framewidth=0.75, frameheight=2.75](0.625,0){ED}{LD}

\ldotsnode[angle=90](0.15,-0.3){dots}
\ldotsnode[angle=90](1.1,-0.3){dots}

\dotnode[dotscale=0.7](3,0.2){dotInt}
\pscircleop(4.2,0.2){oplus} 
\pssignal[signalsep=0.05](4.2,1){ditherOplus}{$z_k$}

\psarc[style=Arrow]{->}(3,0.2){0.5}{110}{250}
\psblock(6,0.2){LPF}{LMMSE}
\dotnode[dotstyle=square*,dotscale=0.001](7.5,0.2){end}

\psset{style=Arrow}
\ncline{ditherD}{ditherCon} 
\ncline[doubleline=true]{in0}{EDin0} 
\ncline[doubleline=true]{in1}{EDin1}
\ncline[doubleline=true]{inL}{EDinL}
\ncline{EDout0}{stream0} \naput{$a^{(j_1)}_{q,n}$}
\ncline{EDout1}{stream1} \naput{$a^{(j_2)}_{q,n}$}
\ncline{EDoutL}{streamL} \naput{$a^{(j_{\size{\set{J}}})}_{q,n}$}

\ncline{-}{dotInt}{stream0}

\ncline{dotInt}{oplus} \naput{$a^{(\set{J})}_{q,k}$}
\ncline{oplus}{LPF} \naput{$\hat{a}^{(\set{J})}_k$}
\ncline{LPF}{end} \naput{$\hat{x}^{(\set{J})}_n$}

\ncline{ditherOplus}{oplus} \nbput[labelsep=0.05,npos=0.8]{--}

\end{pspicture}
\caption{The DSQ coding scheme. Illustrated on the left is the encoder, which produces $L$ descriptions using oversampling and dithered quantization. Illustrated on the right is the decoder operation when a set $\set{J}=\left\{j_1,j_2,...,j_{\size{\set{J}}}\right\}$ of descriptions is received. LC and LD stand for lossless coding and lossless decoding respectively. The time index $n$ corresponds to the original sampling
rate, while $k$ corresponds to the oversampled rate.
}
\label{fig:encoder}
\vspace{-12pt}
\end{figure*}
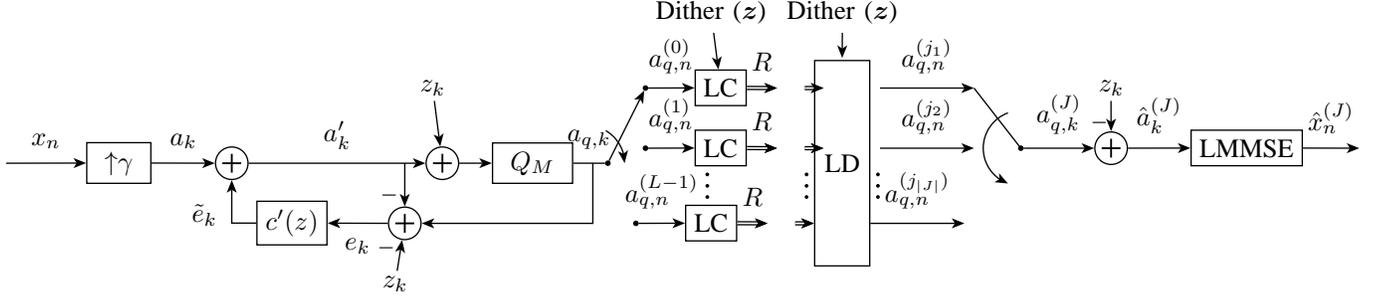

\textbf{DSQ scheme:} Let us now introduce a coding scheme, which generalizes the scheme we proposed in~\cite{Ostergaard&Zamir} and \cite{Mashiach&Ostergaard&Zamir}. This scheme is based on oversampling and entropy-coded dithered (lattice) quantization (ECDQ) with noise shaping at the encoder and linear estimation at the decoder. The oversampling process creates redundant representations of the source, while the noise shaping operation enables controlling the distortions at the different receivers. For simplicity of the exposition, we introduce the scheme using scalar quantization.

The DSQ coding scheme is illustrated in \reffig{fig:encoder}. At the encoder, the source sequence $\seq{x}$ is being oversampled by some oversampling factor $\osr$ to produce the oversampled sequence 
$\seq{a}$\comment{$a_k=\sum_{n=-\infty}^{\infty}x_n \sinc{\frac{k-n\osr}{\osr}}$}, which is bandlimited to $|\omega| \leq \frac{\pi}{\osr}$. Before being quantized, the\comment{ oversampled source} sequence $\seq{a}$ is combined with noise feedback $\tilde{\seq{e}}$, which is created by feeding the quantization error $\seq{e}$ back through a causal filter $C'(z)$. The resulting signal $\seq{a}' = \seq{a} + \tilde{\seq{e}}$ is sequentially quantized using dithered quantizer with second moment $\varE$, to yield the quantized sequence $\seq{a}_q=\seq{Q}(\seq{a'}+\seq{z})$, where the dither $\seq{z}$ is known to both the encoder and the decoder. The dither process is i.i.d., independent of the source and uniformly distributed over a basic cell of the quantizer.

The quantized series $\seq{a}_q$ in the output of the quantizer is being de-multiplexed sample-by-sample to $L$ streams $\{\str{\seqstyle{a}_q}{0}, \str{\seq{a}_q}{1},...,\str{\seq{a}_q}{L-1}\}$, each is losslessly encoded (conditioned on the dither) to yield a description. Since the encoder does not know which descriptions will be received, it encodes each description independently of the others (distributed coding). In Section~\ref{sec:NoRB}, we use sample-by-sample entropy-coding, thus each stream can be losslessly reconstructed by itself, while in Section~\ref{sec:DSQ-RB} we use random binning so that only when receiving at least $K$ descriptions, the corresponding streams can be reconstructed. Since the scheme is time invariant all descriptions have the same rate. We notice that, as in the PPR scheme, the encoding procedure is divided into a lossy step (the quantization) followed by a lossless step.

\begin{figure}
\vspace{8pt}
\begin{center}
\includegraphics{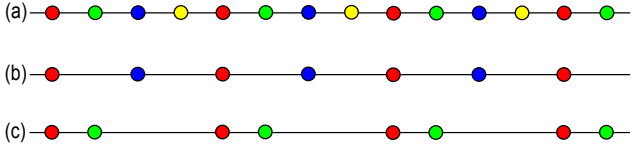}
\caption{A four-description example: (a) all four descriptions; (b) a set of 2 descriptions that correspond to a uniform sampling pattern; and, (c) a set of 2 descriptions that correspond to a non-uniform sampling pattern.}
\label{fig:sampling}
\end{center}
\vspace{-24pt}
\end{figure}

Upon receiving some set of descriptions $\set{J}$, the receiver decodes the $\size{\set{J}}$ streams $\left\{\str{\seq{a}_q}{j}\right\}_{j \in \set{J}}$ and subtracts the dither to get $\big\{\str{\seq{\hat{a}}}{j}\big\}_{j \in \set{J}}$ from which it estimates the source sequence. For simplicity we consider linear estimation, which is asymptotically optimal for Gaussian source and a good lattice quantizer, as the lattice dimension tends to $\infty$ \cite{Ostergaard&Zamir}. Each received description corresponds to noisy uniform sampling of the source at sampling rate of $\osr/L$. The interleaved $\size{\set{J}}$ descriptions at the receiver correspond to either uniform sampling (``uniform receivers'') or periodic \emph{nonuniform} sampling (``nonuniform receivers'') as demonstrated in \reffig{fig:sampling}. For any oversampling factor and noise shaping filter, the reconstruction rule and the resulting distortion may depend not only upon the number of received descriptions but (generally) also upon the descriptions that are received. 

\begin{figure}[!t]
\begin{center}
\begin{tikzpicture}
\newcommand{\x}{3.5}
\newcommand{\y}{1.7}
\newcommand{\xp}{3}
\newcommand{\xl}{1}
\newcommand{\ydc}{0.2}
\newcommand{\yds}{1}

\filldraw[fill=black!20!white, draw=black!50!black]  (-\xp,0) -- (-\xp,\yds) -- (-\xl,\yds) -- (-\xl,\ydc) -- (\xl,\ydc) 
  -- (\xl,\yds) --  (\xp,\yds) 
  -- (\xp,0) ;
\draw[->]  (-\x,0) -- (\x,0);
\draw[->]  (0 ,-0.1) -- (0 ,\y);

\draw (\x ,0) -- (\x,0) node[anchor=north] {$w$};
\draw (\xp,-1pt) -- (\xp,1pt) node[anchor=north] {$\pi$};
\draw (-\xp,-1pt) -- (-\xp,1pt) node[anchor=north] {$-\pi$};
\draw (\xl,-1pt) -- (\xl,1pt) node[anchor=north] {$\pi / \osr$};
\draw (-\xl,-1pt) -- (-\xl,1pt) node[anchor=north] {$-\pi / \osr$};
\draw (0 ,\y) -- (0,\y) node[anchor=east] {$|C(e^{jw})|^2$};

\draw[dotted] (-\xl,0) -- (-\xl,\ydc);
\draw[dotted] (\xl,0) -- (\xl,\ydc);
\draw[dotted] (\xl,\ydc) -- (\x,\ydc) node[anchor=west] {$\delta^{1-\osr}$};
\draw[dotted] (\xp,\yds) -- (\x,\yds) node[anchor=west] {$\delta$};

\end{tikzpicture}
\vspace{-10pt}
\caption{The magnitude spectrum of the optimal noise shaping filter.}
\vspace{1pt}
\label{fig:filter}
\end{center}
\end{figure}
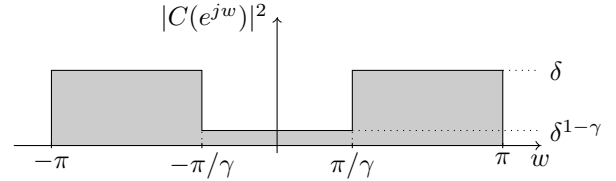

Using the properties of ECDQ (see \cite{Ostergaard&Zamir}) the quantizer output is given by the test channel $\hat{a}_k = a_k+\eps_k$, with ``equivalent noise'' $\eps_k \triangleq \tilde{e}_k +e_k$. $\eps_k$ is statistically independent of the source and is obtained by passing the quantization noise $e_k$ through a monic causal noise-shaping filter $C(z)=C'(z)+1$. Since the quantization error of the dithered quantizer is white with variance $\varE$, it follows that the equivalent noise spectrum is given by $S_{\eps}(w) = |C(e^{jw})|^2\varE$. In this paper we use a monic causal minimum phase filter with a magnitude spectrum that is given in \reffig{fig:filter}. The high-pass nature of the noise-shaping filter causes each couple of descriptions to be negatively correlated, as in Ozarow's test channel (for two descriptions). The parameter $\delta$ controls the shape of quantization noise, and thus the trade-off between $d_K$ and $d_L$. The larger the value of $\delta$ is, the more negatively correlated the descriptions are, and the greater the ratio $d_K/d_L$ is.

While in the discussion above we used scalar quantization, in the next sections we will assume an $M$-dimensional lattice vector quantizer, with a normalized second moment $G_M$. This quantization scheme can be obtained by demultiplexing the original i.i.d. source into $M$ independent parallel processes and applying the scheme in \reffig{fig:encoder} to each, while using one common $M$ dimensional quantizer (see \cite{Ostergaard&Zamir} for details). 

\section{DSQ Scheme with Sub-Nyquist Sampling}
\label{sec:NoRB}
In this section, we study the possibility of achieving the RDF of the Gaussian $K$-or-$L$ problem using DSQ-based scheme with a sub-Nyquist per-description sampling rate. In order to take advantage of the knowledge that at least $K$ descriptions are available at the decoder, each description is sampled at $1/K$ of the source's Nyquist rate. Only when receiving at least $K$ descriptions, the source can be reconstructed at the decoder without aliasing (of the source spectrum).

To achieve the desired sampling rate at each of the descriptions, we use an oversampling factor of $\frac{L}{K}$. When using this oversampling factor, the distortion at the first layer receivers depends not only on the number of received descriptions, but also on the specific set. We begin by considering the uniform receivers and notice that they all have the same distortion. Since uniform sampling when receiving $K$ descriptions is not possible if $K$ does not divide $L$, we assume that $\frac{L}{K}$ is an integer. The following theorem gives an optimality result for this coding scheme for the uniform receivers only.
 
\begin{theorem} \label{theo:KorLuni}
Let $L$ and $K$ be integers such that $K$ divides $L$. The coding rate and distortions at the uniform receivers of the DSQ coding scheme in \reffig{fig:encoder}, with $\osr=\frac{L}{K}$, lattice quantizer of dimension $M$ and the noise shaping filter in  \reffig{fig:filter}, achieve the RDF of the $K$-or-$L$ problem given in \refequ{equ:RKL}, up to a rate loss of at most $\frac{1}{2K}\logtwo{2\pi e G_M}$ bit per source sample. 
\end{theorem}
It is known that there exist lattices where $G_M \to \frac{1}{2\pi e}$ as $M \to \infty$. For such a sequence of good lattice quantizers, the above rate loss tends to zero as $M$ tends to infinity.
\begin{IEEEproof}
Due to lack of space we give a shortened version of the proof, and rely on results from \cite{Mashiach&Ostergaard&Zamir}. By taking a DSQ scheme for $\frac{L}{K}$ descriptions with $\osr=\frac{L}{K}$ and splitting sample-by-sample each of the $\frac{L}{K}$ streams at the ECDQ output (before the lossless encoding) into $K$ streams, we get a $L$-description DSQ scheme. We notice that in the resulting scheme, each set of $K$ descriptions that corresponds to uniform sampling originated from one description of the original scheme. Now, since the original DSQ scheme have the desired properties for the $1$-or-$L/K$ problem by \cite[Theorem 1]{Mashiach&Ostergaard&Zamir} and using the ``scaling property'' of the RDF \refequ{equ:RKL}, we conclude the proof.
\end{IEEEproof}
From \reftheo{theo:KorLuni} we conclude that for the uniform receivers sub-Nyquist sampling is indeed a substitute for random binning. However, the distortion at all other receivers is strictly higher than that at the uniform receivers, a phenomenon known as noise amplification in nonuniform sampling \cite{Mashiach&Zamir}. On the other hand, we notice that when receiving less than $K$ descriptions, while the PPR scheme can not recover the coded descriptions from their bin indices, the DSQ scheme only suffers from an aliasing effect, that decreases as more descriptions are received.



\section{DSQ Scheme with Random Binning}
\label{sec:DSQ-RB}
In this section, in order to avoid noise amplification at the non-uniform receivers, we use an oversampling factor of $L$, thus each description is sampled at the source's Nyquist rate. To use the fact that at least $K$ descriptions are available at the receiver we use random binning as the lossless coding step of the DSQ scheme. We begin by describing the simple reconstruction rules and showing that for oversampling factor of $L$, there is no noise amplification at any of the receivers. Then we prove that when using random binning as the lossless coding step in this DSQ scheme, it has the same performance as the PPR scheme, thus, it is optimal for the $K$-or-$L$ problem. 

\subsection{Avoiding Noise Amplification}
\label{sec:K=1}
We will now present a simple but optimal reconstruction rule from any set $\set{J}$ of received descriptions, and show that the distortion depends only on the number of received descriptions. Each received stream (description) $\str{\hat{\seq{a}}}{i}$ consists of Nyquist rate noisy source samples, taken at sampling times $t_n=n+\frac{i}{L}$. We resample each stream at the source original sampling times $t_n=n$ to yield the ``phase corrected'' stream
\begin{equation}
\tilde{a}^{(i)}_n 
= \sum_{k=-\infty}^{\infty}{\hat{a}^{(i)}_{k}\sinc{n-k-\frac{i}{L}}}
.
\label{equ:phaseCorrectedDef}
\end{equation}
These ``phase corrected'' streams prove useful in evaluating the scheme distortion and rate, using the following lemma.
\begin{lemma} \label{lem:streamsCorr}
For the DSQ coding scheme in \reffig{fig:encoder} with $\osr=L$
and any noise shaping filter $C(w)$ having the same magnitude spectrum as in \reffig{fig:filter}, the ``phase corrected'' streams are given by $\tilde{a}^{(i)}_n = x_n+\tilde{\eps}^{(i)}_n$, where the noise streams $\tilde{\rseq{\eps}}^{(i)}$ are white with variance $\varE \left(\frac{1}{L}\delta^{1-L}+\frac{L-1}{L}\delta\right)$, and the cross-correlation function of $\tilde{\rseq{\eps}}^{(i)}$ and $\tilde{\rseq{\eps}}^{(j)}$ ($i \neq j$) is given by:
\begin{equation}
r_{\tilde{\eps}^{(i)} \tilde{\eps}^{(j)}}(k)=
\left\{ \,
\begin{IEEEeqnarraybox}[][c]{l?s}
\IEEEstrut
-\frac{\varE}{L}(\delta-\delta^{1-L}) , 
 & if $k=0$, \\
0 & if $k\neq 0$
\IEEEstrut
\end{IEEEeqnarraybox}
\right.
\label{equ:Reptilde_ij}
\end{equation}
\end{lemma}

Using \reflem{lem:streamsCorr}, it is clear that optimal linear estimation of the source $\rseq{x}$ from the ``phase corrected'' streams is\comment{ merely their per-sample average multiplied by the Wiener estimation coefficient $\alpha_{\set{J}}$}
\begin{equation}
\hat{\rvar{x}}^{(\set{J})}_n=\frac{\alpha_{\set{J}}}{\size{\set{J}}} \sum_{i\in\set{J}}{\tilde{\rvar{a}}_n^{(i)}},
\label{equ:recJ}
\end{equation}
where $\alpha_{\set{J}} = \varx \left[\varx+\varE\left(\frac{1}{L}\delta^{1-L}+\left(\frac{1}{\size{\set{J}}}-\frac{1}{L}\right)\delta\right)\right]^{-1}$ is the Wiener estimation coefficient.
Thus the reconstruction rule from any set $\set{J}$ of descriptions is a simple two step procedure. In the first step the decoder resample each $\str{\hat{\seq{a}}}{i}$ at the source original sampling times to yield the ``phase corrected'' streams defined in \refequ{equ:phaseCorrectedDef}. Then the reconstruction \refequ{equ:recJ} is merely their per-sample average multiplied by $\alpha_{\set{J}}$. The distortion achieved when using this reconstruction rule is
\begin{equation}
d_{\set{J}}
=\frac{\varx\varE\left(\frac{1}{L}\delta^{1-L}+\left(\frac{1}{\size{\set{J}}}-\frac{1}{L}\right)\delta\right)}
{\varx+\varE\left(\frac{1}{L}\delta^{1-L}+\left(\frac{1}{\size{\set{J}}}-\frac{1}{L}\right)\delta\right)}
,
\label{equ:Dj}
\end{equation}
which depends only on the number of received descriptions. Thus for oversampling factor of $L$ there is no noise amplification (for \emph{any} number of descriptions).

\begin{remark}
For the uniform receivers it can be shown that the reconstruction rule \refequ{equ:recJ} is equivalent to applying a low-pass filter with bandwidth $\frac{\size{\set{J}}\pi}{L}$ and down-sampling by $\size{\set{J}}$.
\end{remark}

\subsection{Achieving the $K$-or-$L$ RDF}
We begin by describing how to apply the random binning encoding to the DSQ scheme. Each of the $L$ quantized streams $\str{\seq{a}_q}{j}$ is divided into blocks of size $N$, which have some distribution over the lattice $\lat^N$ that depends on the dither. Each point of $\lat^{N}$ is randomly and independently assigned to one of $2^{NR}$ bins, where $R$ is the resulting description rate. The \ith description is the bin index of the corresponding stream vector of length $N$. When receiving a set $\set{J}$ of descriptions, the decoder looks for $\size{\set{J}}$ vectors of $\lat^N$, one from each received bin, that are jointly typical given the known dither.\comment{ We notice that the dither, although known at the encoder, has no role in the binning encoding and affects only the decoding procedure.} The next theorem gives an optimality result for this scheme. 

\begin{theorem}\label{theo:Delta-Sigma-Binning}
The $L$-description DSQ coding scheme in \reffig{fig:encoder}, with $\osr=L$, noise shaping filter as in \reffig{fig:filter} and random binning encoding of the quantizer outputs, achieves the RDF of the $K$-or-$L$ problem given in \refequ{equ:RKL}, up to a vanishing ($M \to \infty$) rate loss of at most $\frac{1}{2}\logtwo{2\pi e G_M}$ bit per source sample.
\end{theorem}

\begin{IEEEproof}
Due to space limitations we present only a sketch of the proof. For simplicity of exposition we will use scalar quantization in the derivation, and then extend to lattice quantization. Moreover, we will assume infinite order filter, while it can be shown that the same result holds for $p \to \infty$. 

By \reflem{lem:streamsCorr}, the random vectors $\left(\rvar{x}_n,\tilde{\rvar{a}}^{(0)}_n,...,\tilde{\rvar{a}}^{(L-1)}_n\right)$ and $\left(\rvar{x},\rvar{y}_1,...,\rvar{y}_L\right)$  have the same second moments (and asymptotically the same distribution), where the connection between the PPR's $(\var{V}, \rho)$ and the DSQ's $(\varE,\delta)$ is given by
\begin{equation}
\var{V}=\frac{\varE}{L}\left(\delta^{1-L}+(L-1)\delta\right)
,\;\;\:
\rho\var{V}=-\frac{\varE}{L}(\delta-\delta^{1-L}).
\label{equ:connections}
\end{equation}
Using \refequ{equ:connections}, the DSQ distortions \refequ{equ:Dj} equals the PPR distortions \refequ{equ:PPR_Dj}, when receiving any set of descriptions $\set{J}$ ($\size{\set{J}} \geq K$). 

The DSQ coding rate should be high enough so that for any received set $\set{J}$ of descriptions (bins) such that $\size{\set{J}} \geq K$, there will be only one $\size{\set{J}}$-tuple of typical codewords in the $\size{\set{J}}$ corresponding bins (the one that was sent). We notice that the coding rate is restricted by the ``worst'' (maximum entropy) set $\set{J}$. Now since the input to the random-binning encoder is ergodic, by using \cite[Theorem 2]{cover:ergodicSW} and conditioning on the dither, the scheme's rate is (as $N \to \infty$) 
\begin{equation}
R_{DSQ-RB}=\max_{\set{J}:\size{\set{J}}\geq K} {\EntRate{\str{\rvec{A}_q}{\set{J}}|\str{\rvec{Z}}{\set{J}}}}
,
\label{equ:R_DSQ_max}
\end{equation}
where $\str{\rvec{A}_q}{\set{J}}$ and $\str{\rvec{Z}}{\set{J}}$ are stochastic WSS vector processes. The conditional entropy in \refequ{equ:R_DSQ_max} was calculated in \cite{ZamirKochmanErez} (Appendix A) for ECDQ with feedback.\comment{ While in our case only a sub-sequence of the quantizer output is considered, it can be shown that the same result holds.} Using this result, the causality of the filter and the quantization noise properties
\begin{equation}
\EntRate{\str{\rvec{A}_q}{\set{J}}|\str{\rvec{Z}}{\set{J}}}
=
\DifEntRate{\str{\rvec{\hat{A}}}{\set{J}}}-\DifEnt{\rvar{E}}
.
\label{equ:DSQ_rate_derivation}
\end{equation}
By upper bounding the first expression by the entropy rate of Gaussian process with the same correlation matrix, applying an all-phase filters to correct the phase of each of the streams as in \refequ{equ:phaseCorrectedDef} and using \reflem{lem:streamsCorr} we have
\begin{equation}
\DifEntRate{\str{\rvec{\hat{A}}}{\set{J}}} \leq  \DifEnt{\rvar{y}_1,...,\rvar{y}_K}
,
\label{equ:correctedEnt}
\end{equation}
where the random variables $\rvar{y}_1,...,\rvar{y}_K$ are jointly Gaussian with the same covariance matrix as $\left\{\tilde{\rvar{a}}_n^{(j)}\right\}_{j \in \set{J}}$. Using the properties of the dithered quantization noise we have $\DifEnt{\rvar{E}} = \frac{1}{2}\logtwo{2\pi e \varE} - \frac{1}{2}\logtwo{\frac{2\pi e}{12}} $. Now by using \refequ{equ:Reptilde_ij} and \refequ{equ:connections} we can show that $\DifEnt{\rvar{y}_1,...,\rvar{y}_K}-\frac{1}{2}\logtwo{2\pi e \varE}$ equals the PPR rate ($R_{PPR}$) given in \refequ{equ:R_PPR}, thus
\begin{equation}
|R_{DSQ-RB}-R_{PPR}| \leq  \frac{1}{2}\logtwo{\frac{2\pi e}{12}}
.
\label{equ:DSQ_PPR_Rate_Diff}
\end{equation} 
For $M$ dimensional lattice quantization \refequ{equ:DSQ_PPR_Rate_Diff} still holds when replacing $1/12$ with $G_M$. For a sequence of good lattice quantizers $G_M\to\frac{1}{2\pi e}$, and by the optimality of the PPR scheme for the $K$-or-$L$ problem we conclude the proof.
\end{IEEEproof}
\begin{remark}
We conclude from the proof of \reftheo{theo:Delta-Sigma-Binning} that the proposed scheme achieves the same performance as the PPR scheme, not only for $\size{\set{J}}=K$ but for receiving any $\size{\set{J}}\geq K$.
\end{remark}

\section{Conclusion}
\label{sec:conclusion}
We considered the use of sub-Nyquist sampling as a low-complexity substitute for random binning in symmetric multiple description coding. We conclude that although both random binning and sampling are lossless operations (when receiving at least $K$ descriptions), they have a different impact on the distortion. While random binning does not affect the distortion at all, when using sub-Nyquist sampling some receivers suffer from distortion amplification due to nonuniform sampling. This loss can be avoided in a hybrid coding scheme, which combines Nyquist-sampled DSQ and random binning. 


\section*{Acknowledgment}
This research was supported in part by ISF grant number 870/11 by the Israeli Academy of Science. 

\end{document}